\pdfoutput=1
\def\year{2018}\relax
\documentclass[letterpaper]{article} 
\usepackage{aaai18}  
\usepackage{times}  
\usepackage{helvet}  
\usepackage{courier}  
\usepackage[hyphens]{url}  
\usepackage{graphicx}  
\usepackage{amsmath}
\usepackage{wasysym}
\newcommand{\stitle}[1]{\smallskip\noindent\textbf{#1:\xspace}}

\DeclareMathOperator*{\argmin}{arg\,min}

\newif\ifnotes
\notestrue

\newcommand{\eat}[1]{}

\usepackage{eurosym}
\usepackage{framed}
\usepackage{kpfonts}
\usepackage{xspace}
\usepackage{colortbl}
\usepackage[T1]{fontenc}
\usepackage{fancyvrb}
\usepackage[bottom]{footmisc}
\usepackage{mathtools}
\usepackage{listings}
\usepackage{enumitem}
\usepackage{accents}
\usepackage{booktabs}
\usepackage{graphicx}
\usepackage{caption}
\usepackage{subcaption}
\usepackage{fancyvrb}
\usepackage{color}
\usepackage{balance}
\usepackage[]{algorithm2e}
\usepackage[hidelinks]{hyperref}
\hypersetup{breaklinks=true}

\makeatletter
\def\@copyrightspace{\relax}
\makeatother

\def\pprw{8.5in}
\def\pprh{11in}

\definecolor{mred}{rgb}{.80,.12,.30}
\definecolor{grey}{rgb}{0.5,0.5,0.5}
\definecolor{Purple}{rgb}{.75,0,.85}
\definecolor{lightgray}{gray}{0.95}
\definecolor{mid-gray}{gray}{0.85}
\definecolor{darkred}{rgb}{0.7,0.25,0.25}
\definecolor{darkgreen}{rgb}{0.15,0.55,0.15}
\definecolor{darkblue}{rgb}{0.1,0.1,0.5}
\definecolor{blue}{rgb}{0.2,0.58,.9}
\definecolor{black}{rgb}{0,0,0}

\newcommand{\black}[1]{\textcolor{black}{#1}}
\newcommand{\red}[1]{\textcolor{red}{#1}}

\newcommand{\blue}[1]{\textcolor{blue}{#1}}

\newcommand{\gray}[1]{\textcolor{grey}{#1}}

\begin{document}

\title{PopFactor: Live-Streamer Behavior and Popularity }
\author{Robert Netzorg, Lauren Arnett, Augustin Chaintreau, Eugene Wu \vspace{.35cm}\\ Columbia University in the City of New York \vspace{.35cm}\\ September 2018}

\date{September 2018}

\maketitle

\begin{abstract}
Live video-streaming platforms such as Twitch enable top content creators to reap significant profits and influence.
To that effect, various behavioral norms are recommended to new entrants and those seeking to increase their popularity and success.
Chiefly among them are to simply put in the effort and promote on social media outlets such as Twitter, Instagram, and the like.  
But does following these behaviors indeed have a relationship with eventual popularity?

In this paper, we collect a corpus of Twitch streamer popularity measures---spanning social and financial measures---and their behavior data on Twitch and third party platform.  We also compile a set of community-defined behavioral norms.  We then perform temporal analysis to identify the increased predictive value that a streamer's future behavior contributes to predicting future popularity.   At the population level, we find that behavioral information improves the prediction of relative growth that exceeds the median streamer.  At the individual level, we find that although it is difficult to quickly become successful in absolute terms, streamers that put in considerable effort are more successful than the rest, and that creating social media accounts to promote oneself is effective irrespective of when the accounts are created.  Ultimately, we find that studying the popularity and success of content creators in the long term is a promising and rich research area.
\end{abstract}

\section{Introduction}\label{s:intro}

Live-streaming platforms have recently grown to be of tremendous interest
around the world.  One example is
Twitch\footnote{\small\url{http://www.twitch.com}}, a popular U.S.-based
live-streaming platform focused on video game streaming.  Content creators,
called streamers, create channels to broadcast live videos of themselves
playing video games to multitudes of interested followers.  Top video game
streaming channels on the Twitch platform have been viewed over 1 billion times, and the website attracts over 15M viewers per day. In 2014, Twitch was purchased by Amazon for \$970M.  

A key aspect of live-streaming platforms is that there are tremendous social and financial incentives to grow in popularity.  Viewers can support a particular streamer in a variety of ways. Users can follow a streamer and be notified when the streamer starts a broadcast.  Streamers that have on average $\ge3$ concurrent viewers per broadcast gain additional benefits: users can pay for monthly subscriptions to gain exclusive access to additional content and social features, or spend $\$0.01$ to {\it cheer} on a streamer during a broadcast. Further, the audience can directly donate to a streamer on third-party platforms such as Patreon and TipeeeStream; in 2016, over half a million Twitch viewers donated a total of \$80M to their favorite streamers. When combined with product sponsorships, the top streamers can make upwards of \$4M per year.   
In short, Twitch---along with comparable platforms in the U.S. (e.g., Youtube Live) and around the world (e.g., Meitu, Chushou)---are hugely popular and emerging as large economic forces.  

To this end, the Twitch streamer community has curated a set of behavioral norms for how new streamers can quickly grow their audience.  These behaviors vary from the variety of games to play and how long to stream, to the ways in which streamers can promote their channel on third-party social networks such as Twitter, Instagram, and others.  While there is ample official~\cite{twitchadvice} and unofficial~\cite{twitchreddit,twitchguide} advice how streamers can alter their behavior to become more successful, it is unclear whether or not different behavior indeed affects success, and, if so, whether all of the recommended advice applies equally.  

To supplement this uncertainty, there are an increasing number of articles in the popular media that describe the difficulty of growing a successful streaming career~\cite{redditfulltimetwitch,twitchnoone}, and the considerable work it takes to maintain such a career~\cite{getrichnewyorker,youtubeburn,iceposeidon}.  However, there is a lack of quantitative analysis for how content creator behavior is related with their short term and longer term growth in popularity.  Understanding this dynamic could provide a basis for guidelines about how new streamers should choose to focus their time and resources, and for these platforms to develop tools to aid their content providers.

Towards this goal, we have collected a corpus of Twitch streamers that joined Twitch in 2016, and actively broadcast over the course of two years.  This data contains the streamer activity on Twitch\footnote{The Twitch-specific data was provided by the Twitch data science team.} as well as third-party social media platforms such as Twitter and YouTube.  It also contains several popularity measures that represent general social popularity (number of followers), active popularity (number of concurrent viewers during a broadcast), and financial popularity (number of cheers).    In addition, we surveyed and categorized community recommended behaviors into 6 groups of ``rules'' that are believed to aid in streamer success (e.g., produce more content, promote on Twitter).  

Based on these two datasets, this paper studies the temporal dynamics of Twitch streamer popularity growth during their first year of broadcasting---where the primary growth in popularity occurs.  We seek to understand {\it whether} following behavioral rules established by the Twitch community is related to increased popularity (as defined by different popularity measures), and if so, {\it how}.  To this end, we model this as an inference task, where given a streamer's information  at time $t$, to predict the streamer's popularity at a future time $t+\delta$.  

Simply formulating the prediction task is challenging---a naive model that merely predicts future success would confound past streamer behaviors and success with future behavior that the streamer has control over.   We instead propose to analyze the {\it difference} in predictive power between a baseline model that {\it only} uses past streamer information (e.g., before or at time $t$), with a behavioral model that augments the baseline with behavior information between $t$ and $t+\delta$.  The predictive power gained from adding the latter information is a strong indicator of future behavioral effects\footnote{Note that we do not imply causal relationships between observed streamer behaviors and eventual success, due to possible confounders not present in the dataset.  This is why we report and emphasize the difference in predictive accuracy. }. 

Using this procedure, we analyze behavioral effects on predicting future popularity in absolute terms (ranking in the top-10\% of a measure), as well as predicting future relative growth that is higher than the median growth (Section~\ref{s:prediction}).  In other words, the fast-growing streamers.   We find that across the three popularity measures, future behavior does not contribute to more accurately predicting future popularity in absolute terms.  However, future behavior does contribute to predicting future relative growth, which is arguably important for an individual streamer that is deciding how to rapidly grow her audience over the next several months.   We also find that it is simply very difficult to identify streamers whose financial success will rapidly grow in either the short or long term.    

We also study how streamers may individually grow.  We find that it is very difficult to predict streamers that can grow at a rate to reach Twitch Partner status after 2 years (100 average concurrent viewers).  In contrast to popular media~\cite{twitchnoone}, we find that few streamers broadcast consistently to an empty audience, and that streamers that broadcast as a full-time job ($\ge40$hrs/week) are considerably more successful than the rest. In addition, any time is a good time to start publicizing on third-party social media accounts.   Overall, we find that studying content creator behavior as a predictor of future popularity growth is a promising and impactful research direction and discuss future directions in Section~\ref{s:conclusion}.
\section{Related Work}\label{s:related}

Popularity has been broadly studied across many disciplines, including business, marketing, and social networks.  Here, we survey relevant work on quantitative analysis and prediction of social media popularity.

Much prior work has studied content features that lead to social network virality.  For instance, models to predict Facebook photo re-shares~\cite{cheng2014can}, Twitter re-tweets~\cite{hong2011predicting}, Twitter hashtag usage~\cite{ma2013predicting}, Digg story up-votes~\cite{szabo2010predicting}, and hourly volume of news phrases~\cite{yang2010modeling}. This area of work identifies both content-specific features (e.g., of a potential Tweet), as well as user-specific features (e.g., their popularity, network characteristics), that are predictive of the content's eventual popularity (i.e \# of Shares).  Although these studies may use user popularity as a predictive feature, it is unclear {\it how the user became popular}.  In contrast, we specifically investigate which community-accepted behaviors are predictive of popularity growth over time.

Social network user popularity has primarily focused on predicting a given user's popularity based on network characteristics, and whether such popularity measures are indicative of influence.  For instance, popular Instagram users exhibit broader topical interests~\cite{ferrara2014online}, form reciprocal relationships with other users, and often share common followers~\cite{Kim2017}.  Similarly, popular Twitter users tend of create more original tweets, and retweet less~\cite{fu2016online}.    Other studies have employed information diffusion models~\cite{yang2010modeling} to measure the extent to which popularity results in network-level influence, and found that popular users on Twitter are not the top influencers of hashtag propagation on Twitter.  Ultimately, these studies focus on users that are already popular.  

There has been recent work studying the Twitch ecosystem to understand the intrinsic motivations of Twitch streamers and viewers, and how streamers adopt personas to fit the live-streaming medium.  Others have studied the high volume of chat content that forms during streamer broadcasts, and their community characteristics~\cite{HamiltonPlay}.   In terms of popularity, Kaytoue et al.~\cite{Kaytoue2012} predict viewership dynamics within a given broadcasting session.  Our work builds on this body of research by proposing predictive models of streamer popularity growth based on streamer behaviors.

Unlike predicting the popularity of content, which focuses on predicting popularity in the near future, our goal is to study the process of \emph{becoming popular} on a social network by observing behavioral characteristics over long spans of time.   To the best of our knowledge, there are many community-based anecdotes about effective behavior, and relatively few quantitative or longitudinal studies.  Cha et al. suggest that behavior may be a factor in growing social network influence; Hutto et al. find that how Twitter users interact with their social network affects their follower count~\cite{hutto2013longitudinal};  Chang et al. find that diverse content can help increase followings on Pinterest~\cite{chang2014specialization}.  We extend these ideas by examining a broad set of behaviors derived from the Twitch community and quantitatively studying their ability to predict future popularity growth for varying time ranges.
\section{Twitch Data and Popularity}\label{s:data}

We collected behavioral and popularity information for Twitch streamers that created accounts throughout 2016 and remained active for a year.  The behavioral data includes their broadcasting activity on Twitch.  We also collected activity on third party platforms such as Twitter, YouTube, and Instagram, if those accounts were linked from a streamer's Twitch profile.  The Twitch-specific data was provided by the Twitch data science team.  This section describes our data collection as well as statistics of our sample population.

\subsection{Data Collection}

This study concerns the population of Twitch streamers who began streaming at
some point in 2016 and continued streaming consistently (at least once every two months) for at least a year. Our dataset consists of ~17,682 users and 4
million broadcasts. As we received this data from
Twitch, no data cleaning was required. Due to the selection process for our dataset, however, we cannot be totally sure that popularity dynamics after the first year of streaming is what would occur if the streamers consistently continued to stream into the second year. Therefore, our analysis in Section~\ref{s:prediction} focuses solely on the first year of streaming.

Although Twitch is focused on gamers, it allows non-gaming streamers (e.g., cooking) . Only 1.5\% of all broadcasts in our corpus were non-gaming related, and we do not find that they bias our results.  Thus, we keep them in the dataset. 

\stitle{Third-Party Social Media Data} 
To more completely understand a streamer's presence on the internet, we use links on streamers' profiles to other social media accounts and scrape data that is publicly available from Twitter, YouTube, and Instagram. These three platforms provide temporal insight into a streamer's behavior on external social media accounts---for instance, whether the streamer advertises an upcoming broadcast on Twitter.  It also allows us to study how the streamer's follower community has developed on other platforms.  Using the Twitter, YouTube, and Instagram APIs, we collected the entire posting history for streamers with third-party accounts.  While there are other third-party platforms, such as Patreon or Snapchat, that streamers also link in their profiles, those platforms do not provide access to historical information, so we did not collect data from them.  We also did not collect data from platforms that some Twitch streamers use, such as Discord, TippeeeStream, and Facebook groups because they do not provide public APIs.

\subsection{Popularity Measures} 

Content creators may have different motivations for broadcasting on Twitch~\cite{maslow1943theory,weiner1972theories}---it may be for financial gain, to seek popularity and fame, for social interaction, or because they simply enjoy it.   For this reason, we studied multiple measures of streamer popularity related to total popularity, active viewership, and financial gain. \texttt{Follower} counts measure the number of Twitch users that want to be notified when a streamer begins a new broadcast; concurrent viewer counts (\texttt{Conc. Viewer}) measure the average number of users that watch a streamer's broadcasts for at least a few minutes; cumulative viewer counts (\texttt{Cum. Viewer}) measure the total number of users that watched a streamer's broadcasts in a month (for any amount of time); and \texttt{Cheers} measure the number of $\$0.01$ donations during a streamer's broadcasts.  

\begin{figure}[h]
\centering
\includegraphics[width=.9\columnwidth]{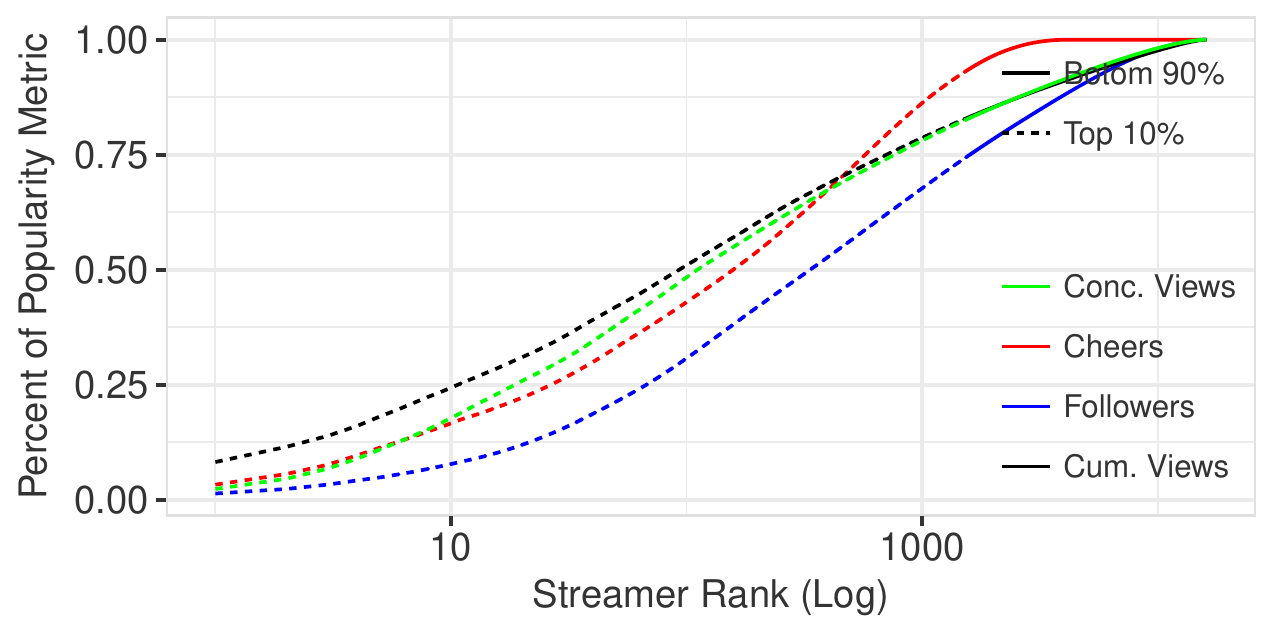}
\caption{Popularity of streamers is highly skewed.  The top 10\% for each measure is dashed.}
\label{f:popdist}
\end{figure}

Figure~\ref{f:popdist} plots the cumulative popularity of top streamers for each popularity measure one year after creating their account on Twitch.  The dashed portion of the line represents the top 0-10\% of streamers for each popularity measure. For instance, the top 10\% of streamers account for 72\% of the total number of followers at the end of one year. Other popularity measures are even more skewed with the top 10\% receiving above 80\% of views and 90\% of cheers. Indeed, nearly 45\%  of streamers never receive a single cheer after one year. The curves for concurrent viewers and cumulative views are nearly identical, indicating that after one year, ranking streamers based on historical popularity or recent viewership audience does not affect the distribution. However, in subsequent sections we will find that the growth dynamics on a per-streamer basis for these two measures are quite different.

\section{From Behavioral Norms to Features}\label{s:rules}

To help new streamers grow their followers, the Twitch community has curated effective behavioral norms into a set of ``rules'' believed to be indicative of popularity growth. We describe our categorization of these rules and how we translate them into features used in our prediction models.

\subsection{Community-Recognized Behaviors}

We surveyed the popular Twitch subreddit\footnote{\small\url{https://www.reddit.com/r/Twitch/}} and community-developed guides~\cite{twitchguide,twitchpopularitybook}.  We then classified them into six general {\it rules}.  For space constraints, we summarize each rule in terms that are not Twitch-specific, and describe the features we extracted to represent each rule.  A comprehensive list of our features and their descriptions is listed in Appendix~\ref{a:features}.

To the best of our ability, we avoided information leakage by excluding behavior features strongly correlated with popularity.  For instance, the audience size during a streaming session and the number of Twitter followers are indicative of popularity. Also, we did not include features based on video stream content (e.g., narration style, emotion) because we lacked access to historical video streams; we leave the analysis of such rules to future work.

\stitle{R1: Produce More Content}
Broadcasting more is considered a core component of becoming popular. Popular streamers tend to stream for 4-8 hours, 5-7 days a week, and new streamers are recommended to to stream $>2-4$ hours per weekday~\cite{twitchguide}.  We computed 4 features measuring number, frequency, and total length of broadcast content.

As an example, Broadcast\_Gap computes the average time between broadcasts, i.e., how long a user is effectively \emph{inactive}. A user will be said to follow the rule ``produce more content'' if she keeps that time small (see Section~\ref{ss:translate} for more details).

\stitle{R2: Release Content Regularly} 
Adhering to a consistent broadcasting schedule is considered a vital part of audience growth.  We computed 2 features that measure whether or not they schedule and to what extent the streamer follows it.

\stitle{R3: Don't Release Overcrowded or Obscure Content}  
Twitch recommends streamers based on popularity, and community wisdom suggests that streamers playing overly popular games will be drowned out by already popular streamers.  On the other hand, overly obscure games will not be interesting to potential followers. We computed 2 features that trace whether or not a streamer plays an overcrowded game and how long they spend playing it.

\stitle{R4: Have a Social Media Presence} 
Linking to, and promoting on, other social media accounts is believed to help build a follower community. We use links on streamer profiles to see whether the streamer has a YouTube or Instagram account. We computed 9 features for third-party social media accounts related to YouTube video and Instagram post metadata.  

\stitle{R5: Twitter is Best for Promotion} 
Twitter is highlighted as one of the best ways to advertise content before and after each broadcast.  We measured general Twitter activity and activity in relation to broadcasts.  We computed 7 features related to Twitter activity and temporal correspondence with broadcasts.  

\stitle{R6: Diversify Your Content}  Based on prior analysis of Pinterest~\cite{chang2014specialization}, diversified content may appeal to a broader audience.   However, streamers typically stream one game, and occasionally mix secondary games.  We computed the number of games played during each broadcast and averaged across each month.

\smallskip
\noindent The scope of our rules was limited by our dataset.  For instance, we did not collect data about the streamer's chat or in-video interactions with the audience during a broadcast~\cite{lurkers}. Despite this, our analysis finds that behavior improves the predictiveness of future follower growth, and we expect the inclusion of additional behavioral rules will strengthen these findings.  

\subsection{Translating Rules into Temporal Features}\label{ss:translate}

We now describe how we distill the question {\it ``did streamer $u$ obey rule $r$ during time interval $[t, t+\delta]$?}'' into a single binary feature that can be used in our prediction model.  We will use Tweet\_Num as the running example; others are simpler or defined similarly. The process requires addressing a number of nuanced challenges.

The first is that the features measured for the above rules are not temporally aligned. Some are per-broadcast while others are per-Tweet.   Second, what does it mean to obey a rule?  Is it relative to the streamer's previous actions, to the rest of the streamer community, or to the streamer that actually succeed?  Third, how to reduce a streamer's rule following, which may change over time, into a single binary value while  losing as little information as possible?  

To address the first challenge, we compute the feature's average value over the time interval $w=[t, t+\delta]$.  For instance, let $B_u = [b_{u,1},\cdots,b_{u,m}]$ be the sequence of $m$ tweets for streamer $u$, and let $b_{u,j}.t$ be the timestamp for the $j^{th}$ tweet. We define $f_{u,w} = count(\{b_{u,j} | b_{u,j}.t \in w\})$ as the number of tweets in the time interval $w$. 

For the second challenge, we observe that the community rules are typically described in relation to the behavior of popular streamers (e.g., broadcast as long as popular streamers) or the streamer community at large (e.g., you should avoid playing the game everyone is playing). For this reason, we interpret obeying a rule as following it {\it more} than the general community, \emph{in a way that imitates popular users}. To best address our third challenge we have to find a binary value that captures as much information as possible. This means that the cutoff $C_f$ for each measure need to be feature specific, where $u$ obeys the rule if her feature is comparing to the cutoff in ways that resemble popular users. Therefore, a user $u$ where $f_{u,w}>C_f$ is given a value of 1 and 0 otherwise. To ensure that this principle is best quantified, the cutoff is chosen as the values $C_f$ where the fraction of streamers following the rules differ the most between popular (top-10\% streamers) and unpopular (bottom-90\% streamers), i.e.,
 \[C_f=\underset{0\leq k\leq1}{argmax}(\left|pop_{k}-unpop_{k}\right|)\]
where $pop_k$ is the fraction of popular streamers whose feature value $f_{u,w}$ is above the $k^{th}$ percentile among all streamers, and $unpop_k$ be defined accordingly for unpopular streamers. Note that, since we will later use a feature to study popularity either in terms of followers, views or cheers, we redefine top-10\% streamer for each case, which means the cutoff will be different (albeit very close) for different prediction tasks.

\begin{figure}[h]
\centering
\includegraphics[width=\columnwidth]{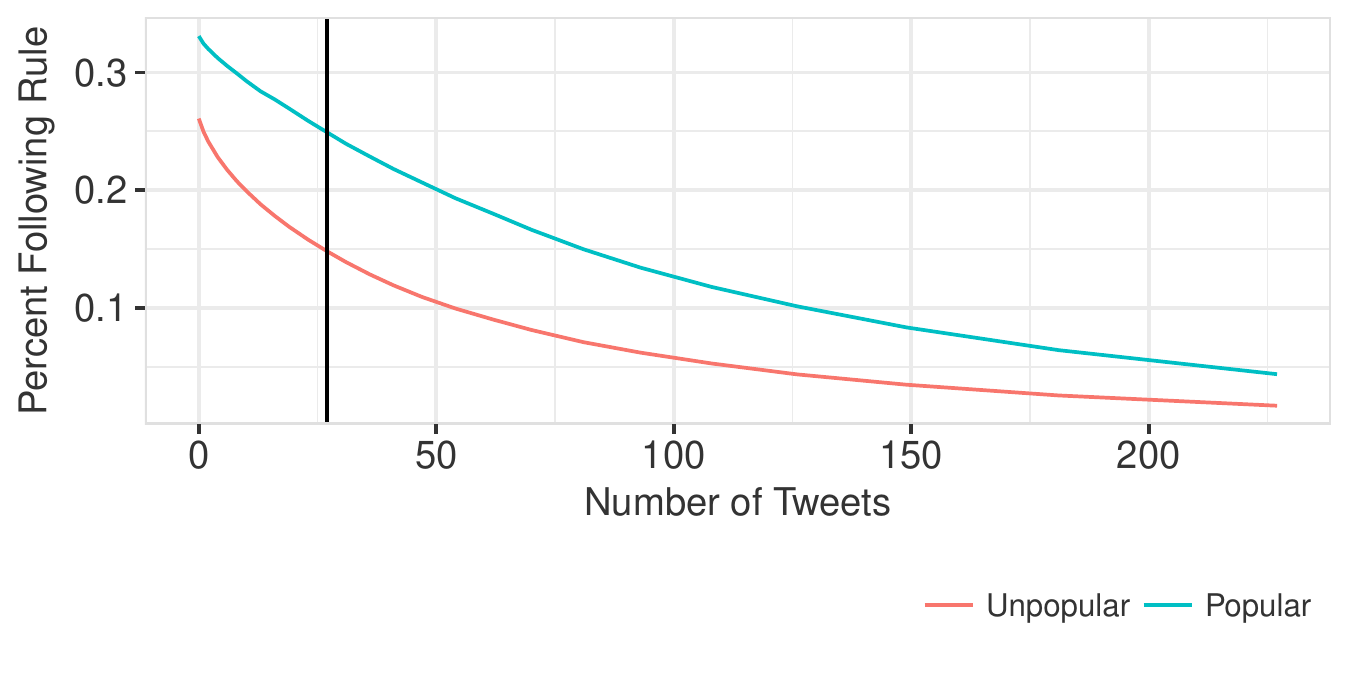}
\caption{Example of feature and cutoff choices for rule ``Tweet More''. The vertical line indicates the dynamic cutoff chosen by our method, while median (0 Tweets) offers less information.}
\label{f:broad_perc}
\end{figure}

An alternative method is to pick cutoffs that ignore popular users and their different behaviors, for instance choose $C_f$ to be the median of $f$ among all streamers. Although for many features it resembles our method\footnote{When running the analysis using the median, even, our results regarding the rules the remain the same, although the feature Twitter Live does become significant for the follower task}, it has two drawbacks: first, in several cases the median (or another arbitrarily chosen percentile) is often degenerate because some features are heavily skewed. Second, when a simple median cutoff exhibits negative results (indicating that behaviors are not adding accuracy in prediction) one can still ask if behavior could possibly help with a more \emph{informative} cutoff. As an example Figure~\ref{f:broad_perc} plots the distribution of Tweet\_Num for the popular (top-10\% streamers in terms of follower count) and unpopular (the other streamers) subpopulations. The median over the whole population (i.e., 0 Tweets) is not as informative to separate popular from unpopular streamers based on that behavior alone. In contrast, the cutoff choice we present avoids that trap unless of course the feature is entirely uninformative.

Using the above procedure, we can compute one feature vector for each user in each time interval for each popularity measure.  In the next section, we define different measures of popular/unpopular and time interval in order to understand the temporal dynamics of streamer behavior on popularity.  

\section{Population-Level Dynamics}\label{s:prediction}

\begin{figure}[h]
  \centering
  \begin{subfigure}[t]{.49\columnwidth}
    \centering
    \includegraphics[width=\columnwidth]{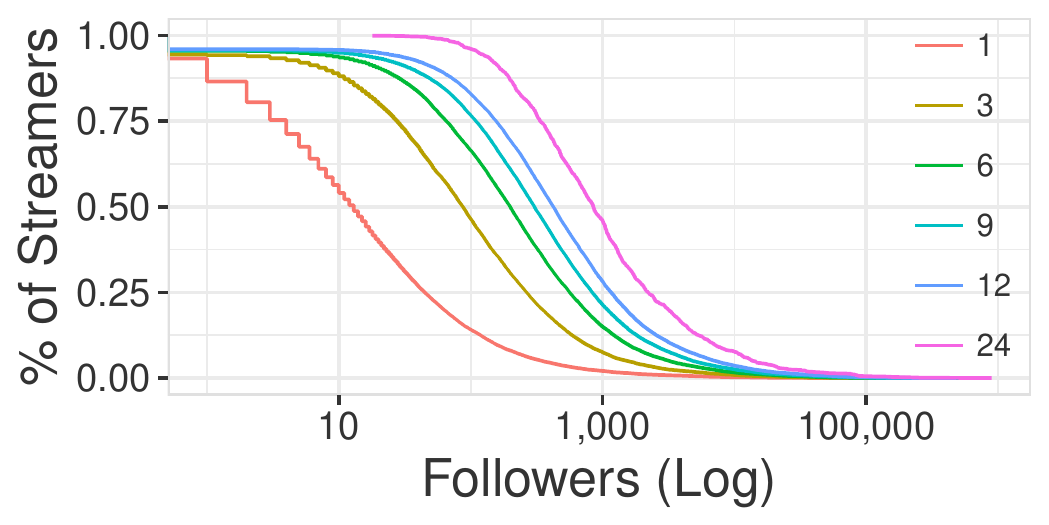}
    \caption{Follower Count.}
    \label{f:ccdf_fols}
  \end{subfigure}
  \begin{subfigure}[t]{.49\columnwidth}
    \centering
    \includegraphics[width=\columnwidth]{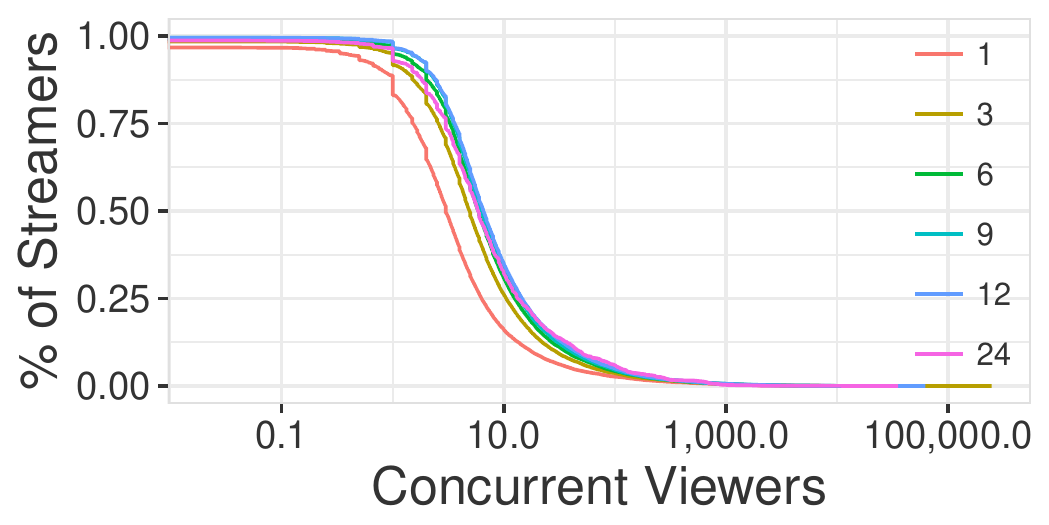}
    \caption{Avg Concurrent Viewers.}
    \label{f:ccdf_views}
  \end{subfigure}\\
  \begin{subfigure}[t]{.49\columnwidth}
    \centering
    \includegraphics[width=\columnwidth]{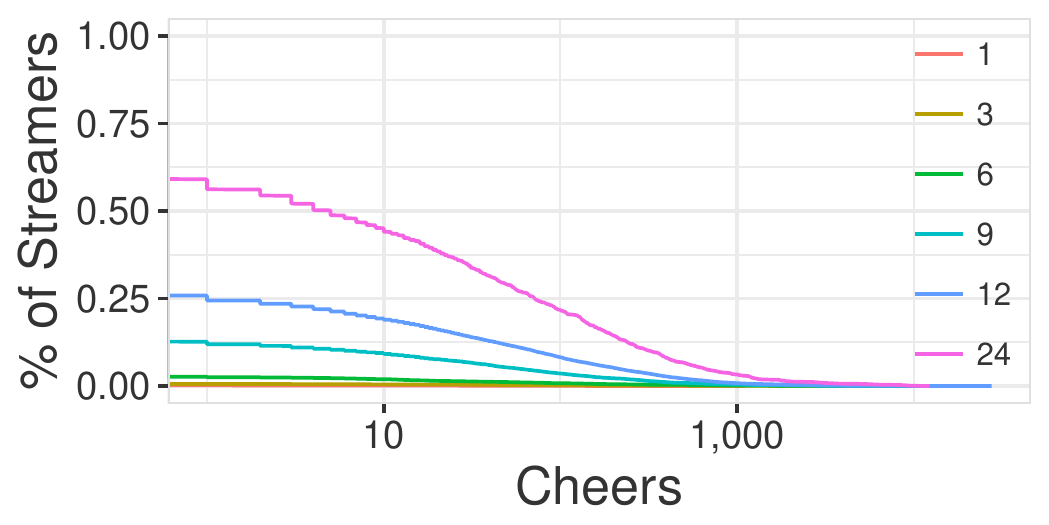}
    \caption{Cheers.}
    \label{f:ccdf_cheers}
  \end{subfigure}
  \begin{subfigure}[t]{.49\columnwidth}
    \centering
    \includegraphics[width=\columnwidth]{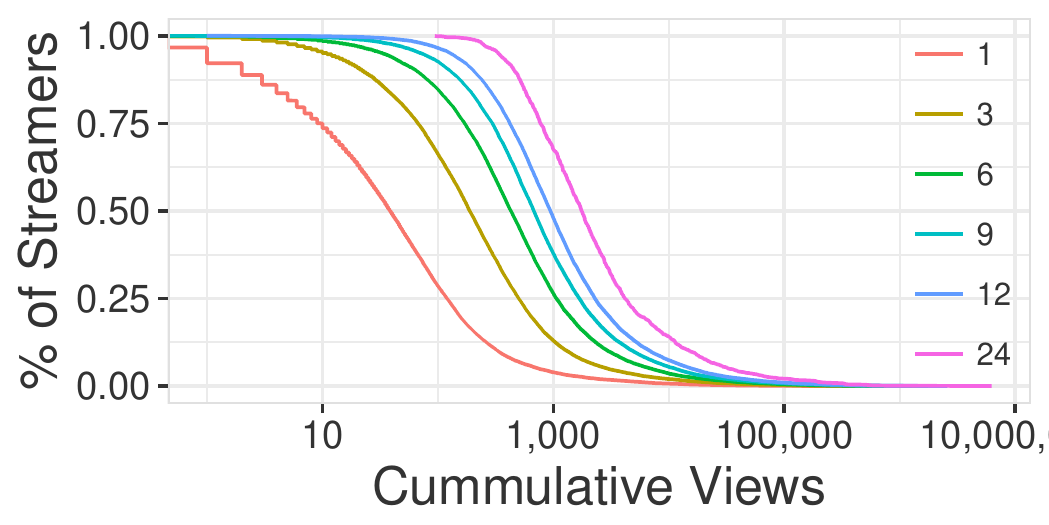}
    \caption{Cumulative Views.}
    \label{f:ccdf_ccu}
  \end{subfigure}
  \caption{\small \% Percentage of streamers (y-axis) with same or more (a) followers, (b) average concurrent viewers, (c) cheers, and (d) cummulative views at different account ages (lines)}
  \label{f:ccdf}
\end{figure}

This section analyzes streamer behaviors that are correlated with the four popularity measures described in Section~\ref{s:data},  We begin with Figure~\ref{f:ccdf}, which shows a population level view of these measures at different months since account creation (streamer age).

For measures such as follower count and cumulative views, growth appears consistent across months for users of all percentiles.  The median user gains 102 and 105 follows in months 1 and 2.   

However, measures such as concurrent viewership and cheers are more elusive.  Even after 2 years, very few streamers reach 100 concurrent viewers, nor more than \$10 in cheers per month (one cheer is 1$\cent$).

The question then becomes: what behaviors dilineate the unpopular streamer from her popular counterpart?  The rest of this section studies this question by using the behavioral factors defined by the community rules and rule-following (Section~\ref{s:rules}).

\subsection{Temporal Analysis Methodology}

The rest of this section studies how behavior at age $t$, measured as the degree of rule-following described above, is correlated with future popularity at age $t+\delta$.    Intuitively, this is challenging because popularity, behavior, and time are intricately connected. In particular, a naive popularity prediction task could confound factors such as current status with behavioral factors.   Without access to randomized experiments and artifacts that can help infer causality, we present a temporal analysis method to minimize the effects of confounding factors and isolate behavioral effects.  

The main idea is to use a strong baseline model $F_{cur}$ that uses all relevant information at time $t$ to predict eventual popularity at $t+\delta$, and compare it with a behavioral model $F_{cur+b}$ that {\it additionally includes behavioral features}.  The difference in predictive accuracy between the two models describes the additional predictive power that behavior accounts for.   We now define the two models.

\stitle{Strong Baseline~($F_{cur}$)} We formulate a binary inference task. The model input includes \emph{all} information on a streamer's popularity and actions, including on third-party platforms, up to age $t$ (e.g., all data prior to age of 4 months). The goal, or output, is to accurately predict whether the streamer was among the top 10\% of a given popularity measure (e.g., top 10\% most followers) by the end of the interval $t+\delta$ (e.g., age of 6 months if $\delta=2$). We call this {\it Absolute Popularity}, as it measures popularity in absolute terms.    

In contrast, an individual streamer may simply care about rapidly growing.  Thus we also define {\it Relative Popularity Growth} by whether or not the streamer's popularity measure increases more than the median streamer's growth.  For instance, if the follower count grew 10\% over 2 months and the median only grew 5\% during the same period, then the streamer had high relative follower count growth and the model should predict $1$.  If not, then the model should predict $0$.  We evaluate both absolute and relative popularity in the following experiments.

Note that $F_{cur}$ carefully accounts for the effect of age.  It uses supervised training to interpolate a growth trajectory using past information until $t$ to estimate the expected outcome at $t+\delta$.  For a fixed age interval size (e.g., $\delta=3$),  we pool the intervals at each monthly starting age (e.g., [1m-4m],[2m-5m],$\cdots$), and report test AUC\footnote{\small Area Under the Curve (AUC) of the ROC Curve describes a model's predictive power and is indifferent to class imbalances, a common problem for using accuracy. 0.5 AUC signifies random guesses, while 1.0 AUC signifies a perfect classifier. The AUC can be interpreted as the probability of correctly ranking a positive and negative class.} using an 80-20 train-test split (performed on the entire dataset before the temporal partition; for each window $[t, t+\delta]$, 20\% of users are held out at random). We use a logistic regression model because the contribution of each behavioral feature can be interpreted by the weights of the model.    

Formally, the prediction task is as follows.  Let $X^t$ and $y^t$ be the set of features and binary popularity outcomes at time $t$ across all users, and let $\delta$ be the time interval size.  The task is to learn a set of linear feature coefficients $A$ that minimize the non-regularized logistic regression:
$$A^* = \argmin_{A} \sum_{t\in [1, 12-\delta]} | Y^{t+\delta} - AX^t |^2_2$$

\stitle{Behavior Model~($F_{cur+b}$)} The behavior model \blue{$F_{cur+b}$} augments the inputs with behavioral features observed \emph{during} the age interval $[t,t+\delta]$ as defined in Section~\ref{ss:translate}.   Since it has more information, it is expected to return a higher AUC. However, note that all past behaviors and popularity of that streamer were previously included, so the {\it only new information} concerns the unexpected/unpredictable changes in the behavior. By focusing on the \emph{AUC gain} $F_{cur+b}-F_{cur}$ rather than absolute model accuracies, we can more confidently isolate how changes in behavior affect the prediction of popularity. Hence, a large difference between the two models is less likely to be due to a pre-existing factor.

 \begin{figure*}[t]
   \centering
   \captionsetup[subfigure]{justification=centering}
   \begin{subfigure}[t]{0.25\paperwidth}
	\includegraphics[width=0.25\paperwidth]{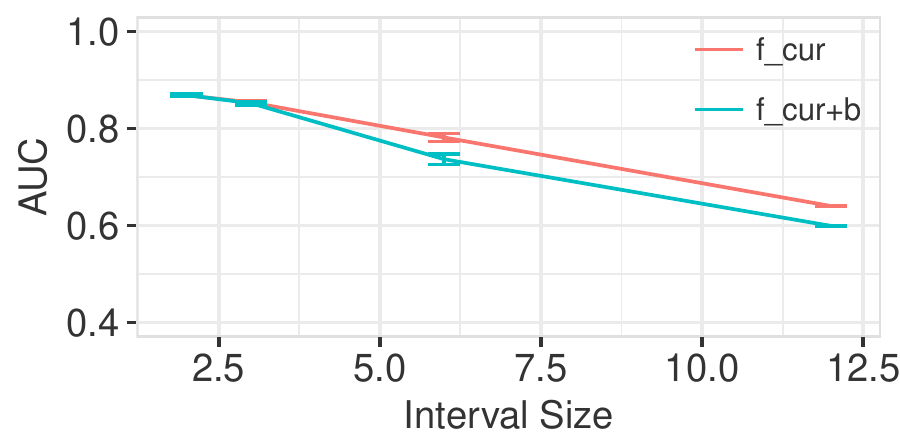}
  \caption{\small Behavior does not contribute to predicting absolute popularity.}
	\label{f:abswindows}
	\end{subfigure}
  \begin{subfigure}[t]{0.25\paperwidth}
	\includegraphics[width=0.25\paperwidth]{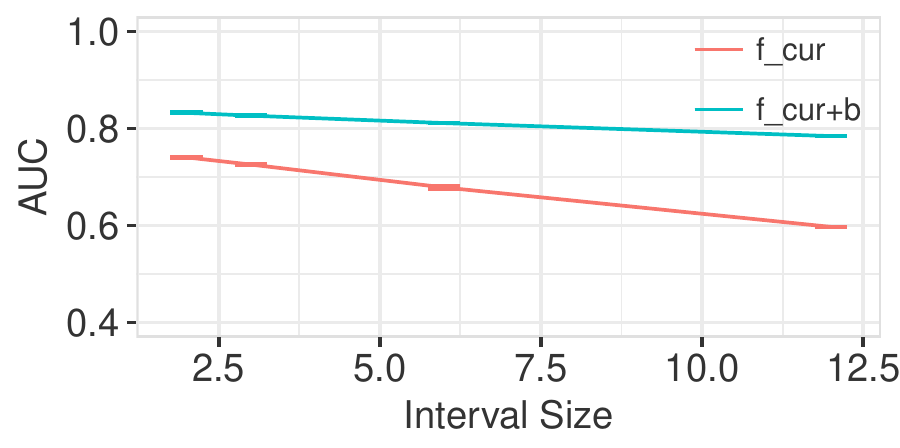}
  \caption{\small Behavior consistently improves prediction of relative follower growth.}
	\label{f:relwindows}
	\end{subfigure}
  \begin{subfigure}[t]{0.25\paperwidth}
	\includegraphics[width=0.25\paperwidth]{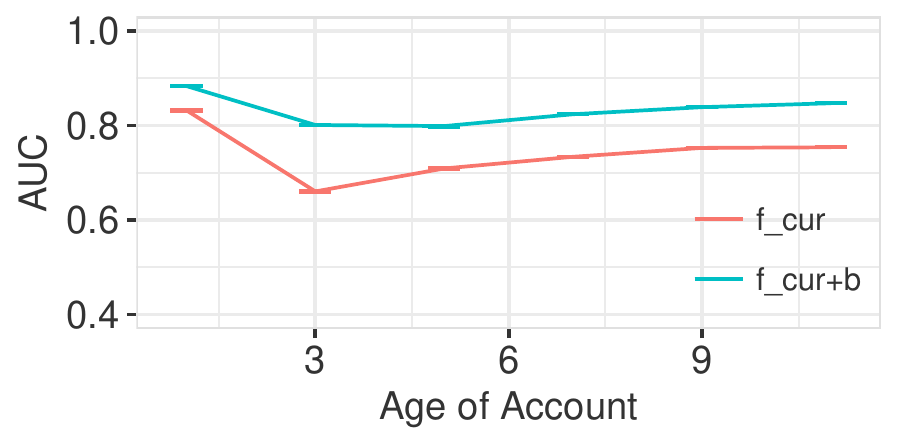}
  \caption{\small Behavior consistently improves relative growth over a 2-month interval, irrespective of streamer age.}
	\label{f:twomonthrel}
	\end{subfigure}
    \caption{\small AUC of \red{\(F_{cur}\)} vs. \blue{\(F_{cur+b}\)}. (a, b) show absolute and relative follower growth for increasing interval sizes (further in the future). (c) shows relative follower growth over 2-month intervals starting at different ages.}
  \label{f:window_follower}
 \end{figure*}

\subsection{Behavior and Follower Growth}
To start, we study the predictiveness of behaviors on absolute and relative popularity in terms of number of followers (Figure~\ref{f:window_follower}). Figure~\ref{f:abswindows} shows that over short time intervals (2 months), the baseline model \red{$F_{cur}$} can predict absolute popularity with nearly 0.87 AUC.  This is because the most popular users typically maintain their status in the short term.  In contrast, over longer periods (1 year), the AUC decreases substantially to as low as $0.65$.   Knowing future behavior (\blue{$F_{cur+b}$}) actually decreases the AUC over the long term to nearly as low as $0.6$, which is slightly above random chance of $0.5$. For the absolute popularity task, prior popularity goes a long way in identifying the future highly popular from the rest (i.e., 55\% of users who are in the top 10\% most followed in the first month end the year in the top 10\%), explaining why behavior would not provide much of a predictive boost.

Figure~\ref{f:relwindows} shows that predicting relative growth using $F_{cur}$ shares a similar trend,  but is generally harder to predict, than absolute popularity.  Incorporating future behavior provides a considerable boost in AUC---by $0.2$ over a 1 year interval.  This is consistent with community expectations that streamer behavior can affect the rate of growth.  More surprisingly, the AUC for $F_{cur+b}$ is almost flat as the time interval increases.  This suggests that behavior may be a strong contributor to a streamer's rate of follower growth over both short and long term---there is potential to control one's popularity in a predictable manner.

To account for streamer age, Figure~\ref{f:twomonthrel} reports the AUC for relative growth, but fixes the interval size to $\delta=2$ months and varies the age at the start of the interval (x-axis).  We find that behavior is indeed important throughout the first year (16\% gain on average), and is highest at 4 months (23\% gain).  We note that the first interval is dramatically higher than the other intervals due to sampling biases.  For example, many professional gamers and previously-popular streamers bring their fans when they create their Twitch account, which exacerbates the distinctions between seemingly high and low growth streamers, making the prediction problem simpler in the first interval.  Interestingly, even then, behavior matters. 

 \begin{figure*}[hbt]
   \centering
   \captionsetup[subfigure]{justification=centering}
   \begin{subfigure}[t]{0.3\textwidth}
	\includegraphics[width=\columnwidth]{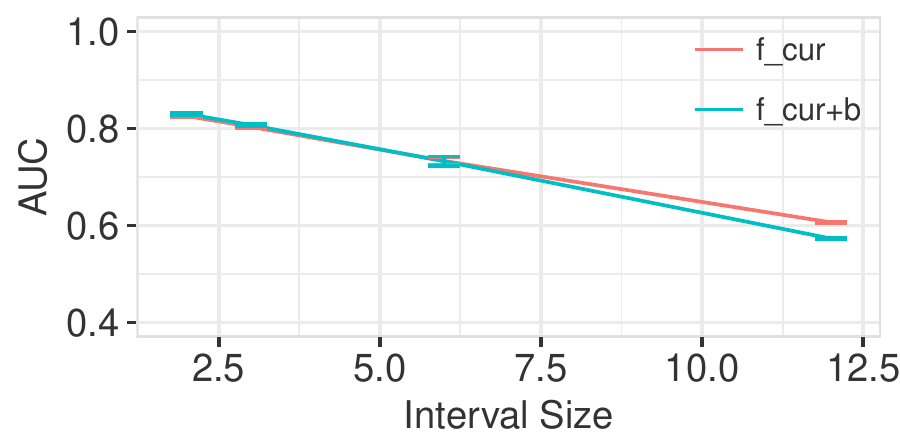}
  \caption{AUC for absolute popularity (average conc. views).}
	\label{f:abs_views}
	\end{subfigure}
  \begin{subfigure}[t]{0.3\textwidth}
	\includegraphics[width=\columnwidth]{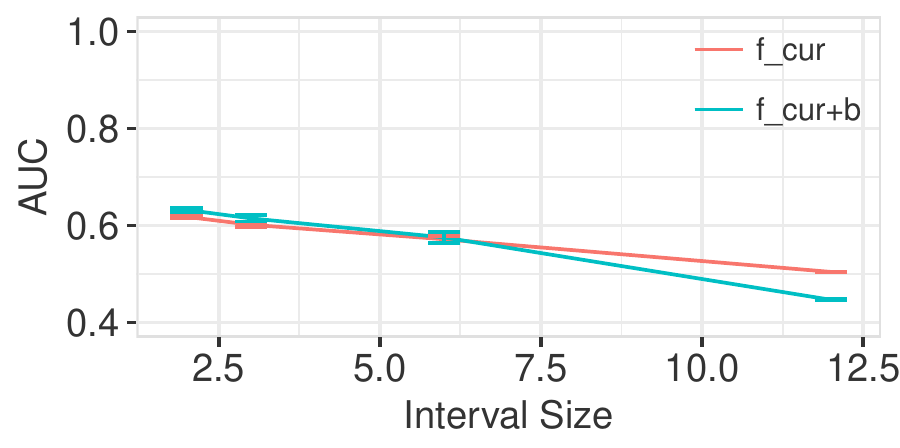}
  \caption{AUC for absolute popularity (cheers).}
	\label{f:abs_cheers}
	\end{subfigure}
  \begin{subfigure}[t]{0.3\textwidth}
	\includegraphics[width=\columnwidth]{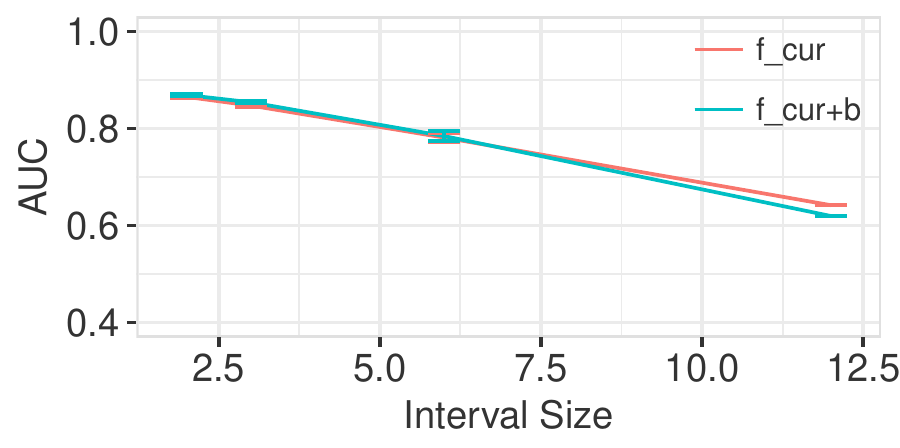}
  \caption{AUC for absolute popularity (cumulative views).}
	\label{f:abs_cumviews}
	\end{subfigure}\\
  \begin{subfigure}[t]{0.3\textwidth}
	\includegraphics[width=\columnwidth]{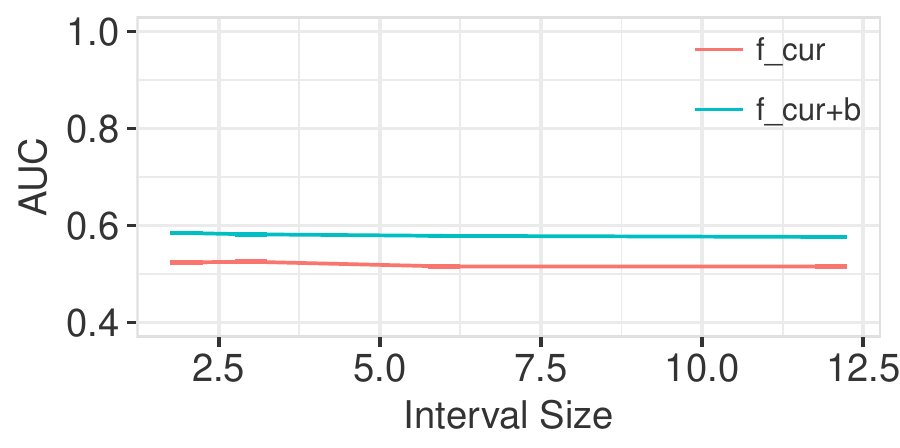}
  \caption{AUC for relative popularity growth (average conc. views).}
	\label{f:rel_views}
	\end{subfigure}
  \begin{subfigure}[t]{0.3\textwidth}
	\includegraphics[width=\columnwidth]{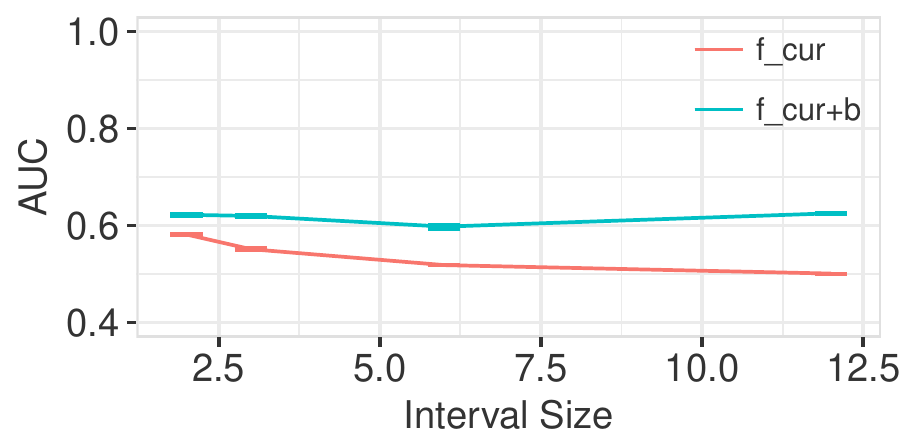}
  \caption{AUC for relative popularity growth (cheers).}
	\label{f:rel_cheers}
	\end{subfigure}
  \begin{subfigure}[t]{0.3\textwidth}
	\includegraphics[width=\columnwidth]{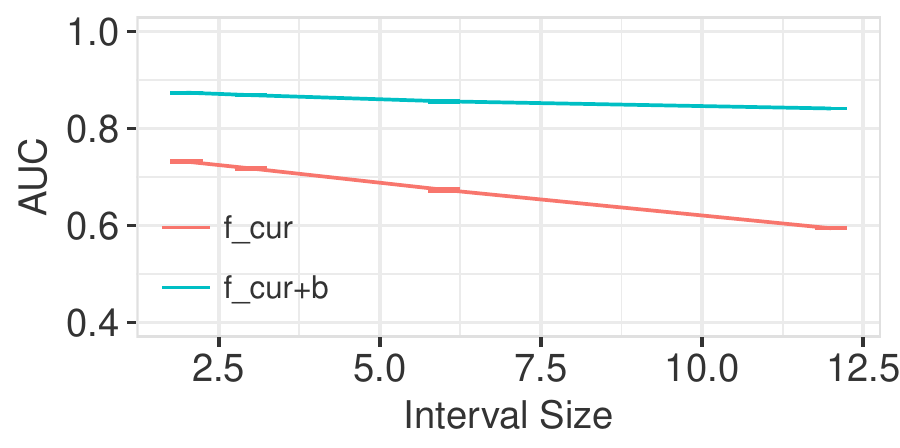}
  \caption{AUC for relative popularity growth (cumulative views).}
	\label{f:rel_cumviews}
	\end{subfigure}
    \caption{\small AUC of \red{\(F_{cur}\)} vs. \blue{\(F_{cur+b}\)} over different interval sizes ($\delta$) to predict (left) average concurrent views, (middle) \# of cheers, and (right) cumulative views.  Error bars denote 1 standard error. }
    \label{f:window_popularity}
 \end{figure*}

\subsection{Additional Popularity Measures}

As discussed in Section~\ref{s:data}, the Twitch ecosystem offers other definitions of popularity beyond follower count. For instance, the average concurrent viewership measures how many users concurrently watch a streamer's average broadcast for longer than a few minutes.  This measure is important because followers may not necessarily watch the streamer.  Many streamers broadcast on Twitch in the hopes of potentially making money, and the number of Cheers is a monetary measure of popularity.  A third measure is the cumulative total views, which measures the total number of times a streamer's broadcasts have been viewed.  This is a cumulative statistic similar to followers, and although it is not used by Twitch, other platforms such as TikTok and YouTube report it. Finally,  views and followers are moderately correlated (0.44), cheers and followers are weakly correlated (0.26), and cumulative views and followers are highly correlated (0.88). 

Unlike follower count, concurrent views and cheers are much more volatile metrics of popularity, and more difficult to attain. After streaming for 2 years, only 55\% of streamers receive a single cheer, 19\% earn \$100, and 4\% earn \$1,000. Further, we note that unlike the other measures, average concurrent viewers does not grow monotonically and can fluctuate considerably from month to month, and broadcast to broadcast.  We find that the difficulty of attaining any concurrent viewers (the median user has $\approx6$ concurrent viewers) impacts the overall accuracy of the predictive models.

\stitle{Absolute Popularity}
The first row of figures in Figure~\ref{f:window_popularity} report the AUC curves for $F_{cur}$ and $F_{cur+b}$ using the absolute popularity of the three measures.  We find that the curves for concurrent viewership and cumulative views are consistent with the followers measure in Figure~\ref{f:abswindows}.  In contrast, there are so few streamers with more than a single cheer that both models perform near randomly, although behavior contributes a slight gain over the 1 year interval.

\stitle{Relative Popularity Growth}
The second row reports AUC curves for relative popularity growth of the three measures.
The cumulative views curves in Figure~\ref{f:rel_cumviews} are nearly identical to the corresponding followers curves in Figure~\ref{f:relwindows}. This similarity makes sense in light of the fact that both measures are monotonic and highly correlated. The prediction ease of $F_{cur+b}$ on cumulative views suggests that, across interval sizes, behavior is indicative of distinguishing between highly and seldom viewed streamers. 

In contrast, the curves for concurrent views and cheers are considerably different.  Both $F_{cur}$ models perform nearly randomly, and although behavior features increases the AUC by nearly 0.1, the overall accuracy is still very low (around 0.6 for both measures).  This suggests that community accepted behavioral rules may not be enough if a streamer is focused on monetary or viewership success.

\begin{table}[hp] \centering \small
\begin{tabular}{@{\extracolsep{5pt}} rlll}
\textbf{} & \textbf{} & \textbf{Cum.} & \textbf{Concur.}\\\textbf{Feature} & \textbf{Followers} & \textbf{Views} & \textbf{Views} \\
\hline \\[-1.8ex]
\# Broadcast & \black{\ 0.85$^{**}$} & \black{\ 1.21$^{**}$} & \black{\ 0.22$^*$}\\
{\gray{Instagram Adv}} & \black{\ 0.77} & \black{\ 0.22} & \black{\ 0.09}\\
{\gray{Tweet After Gap}} & \black{\ 0.72$^{**}$} & \black{\ 0.38$^*$} & \black{\ 0.10}\\
{\gray{YouTube Adv}} & \black{\ 0.61$^*$} & \black{\ 0.50} & \red{\textbf{-0.04}}\\
Broadcast Len & \black{\ 0.58$^{**}$} & \black{\ 0.57$^*$} & \black{\ 0.41$^{**}$}\\
Sched Regularity & \black{\ 0.48$^{**}$} & \black{\ 0.95$^{**}$} & \black{\ 0.22}\\
{\gray{\# Tweet}} & \black{\ 0.37} & \black{\ 0.49} & \black{\ 0.06}\\
{\gray{Tweet Before Gap}} & \black{\ 0.37$^*$} & \black{\ 0.62$^{**}$} & \red{\textbf{-0.09}}\\
\# Days & \black{\ 0.27$^{**}$} & \black{\ 0.60$^{**}$} & \black{\ 0.14}\\
{\gray{\# Twitter Replies}} & \black{\ 0.23} & \black{\ 0.14} & \black{\ 0.08}\\
{\gray{Twitter Live}} & \black{\ 0.20} & \black{\ 0.10} & \black{-0.00}\\
{\gray{Twitter Adv}} & \black{\ 0.19} & \red{\textbf{-0.08}} & \red{\textbf{-0.38}}\\
\# Games & \black{\ 0.16$^*$} & \red{\textbf{-0.06}} & \red{\textbf{-0.07}}\\
{\gray{Instagr. Post Len}} & \black{\ 0.12} & \black{\ 0.11} & \red{\textbf{-0.09}}\\
{\gray{\# Instagram Posts}} & \black{\ 0.12} & \black{\ 0.11} & \black{\ 0.13}\\
{\gray{\# Tags/Instag. Post}} & \black{\ 0.12} & \black{\ 0.11} & \red{\textbf{-0.09}}\\
{\gray{\# YouTube}} & \black{\ 0.01} & \red{\textbf{-0.16}} & \blue{\textbf{\ 0.11}}\\
{\gray{YouTube Title Len}} & \black{\ 0.01} & \red{\textbf{-0.16}} & \blue{\textbf{\ 0.13}}\\
\# Popular Game & \black{-0.00$^{**}$} & \black{\ 0.00$^*$} & \black{-0.00}\\
Gap Btwn Broadcasts & \black{-2.40$^{**}$} & \black{-3.21$^{**}$} & \black{-0.17}\\
{\gray{Tweet Len}} & \black{-0.84$^{**}$} & \black{-0.99$^{**}$} & \black{-0.09}\\
{\gray{YouTube Desc Len}} & \black{-0.14} & \black{-0.04} & \blue{\textbf{\ 0.02}}\\
Unique Games & \black{-0.04} & \black{-0.02} & \black{-0.16$^*$}\\
{\gray{YouTube Video Len}} & \black{-0.01} & \blue{\textbf{\ 0.14}} & \red{\textbf{-0.25}} \\
\end{tabular}
  \caption{\small
     Coefficients of features in the $F_{cur+b}$ model for relative popularity growth over a two month time interval 
    (*: p-value < 0.1, **: p-value<0.05).
    Third-party social media features are \gray{colored in gray}.
    Coefficients that changed from negative in the followers model to positive are colored \blue{blue},
    and colored \red{red} if from positive to negative.
    }
  \label{t:coefs}
\end{table}

\subsection{Comparing Feature Coefficients}

Table~\ref{t:coefs} summarizes each feature's coefficients in the $F_{cur+b}$ relative popularity growth models for each popularity measure; we exclude cheers because the extreme skew of streamers that receive cheers is degenerate and caused the model to perform poorly.  We use $^*$ and $^{**}$ to denote significance at $<0.1$ and $<0.05$ levels.  The p-value for each feature was computed separately by using two-tailed t-test.  For convenience, we summarize how the coefficients change between each pair of models in the final three columns. 

In order to assure the robustness of our coefficient estimates, we ran a correlation analysis to see if our model contained a set of features that were possible colienar with one another. Several features, namely Instagram length, Instagram num, and Tags num, are highly correlated with one another in the follower task. We removed these features and reran to model to find that the significance and coefficient magnitudes remained the same. Because removing the features did not impact the results, we include them here to provide a full analysis of the feature set.

We find that most features have a positive correlation with follower growth.  In particular, regularly broadcasting more often, and for longer periods of time (Broadcast \#, Broadcast Len, Sched Regularity) are all highly correlated with follower growth.  In fact, Broadcast Gap has a very high negative coefficient, which penalizes long periods between consecutive broadcasts.   In addition, advertising on different social media platforms by linking to upcoming streams (Instagram Adv, Youtube Adv, Twitter Adv), and by simply posting (Tweet \#) are highly correlated.  There is a slight negative correlation with longer Tweets and YouTube descriptions.  Thus in general, simply increasing the volume of activity appears to correlate highly with follower growth.  

These results appear similar for the cumulative views model as well.  Although a small number of features, such as Twitter advertisements, the number of games played, and YouTube posts become negatively correlated, the coefficients are not statistically significant. We highlight in \red{red} the features whose coefficients flipped from positive in the followers model to negative, or \blue{blue} if the opposite occured.

The model for concurrent viewers is far more difficult to predict in terms of AUC than the preceeding two measures, and it is also highlighted in the discrepency between its coefficients and the coefficients for the followers model.  For instance, very few features have coefficients that are statistically significant---broadcasting more and longer continue to be the primary features.  Other features, such as advertising on third-party platforms, switch to having no or silghtly negative coefficients.  In fact, most third-party features have negligible coefficients.

All our features tend to predict high growth and not the opposite, according to norms and recommendations of that community. Our results reveal that not all behaviors are as predictive as the community would believe them to be. For all of the behaviors, Table~\ref{t:coefs} indicates that the community was either right or overconfident about the predictiveness of a particular feature, but never so poorly wrong as to say that one feature predicts high growth when it actually predicts low growth. Rules such as Activity, Twitter Promotion, and Regularity seem to hold their weight in terms of importance, but other community-defined rules like Social Media usage or Avoid Playing Popular Games seem to not matter as much towards the growth task. 

\section{Streamer-Centric Analysis}\label{s:investigation}

The previous section studies the relationship between behavior and popularity as compared to the entire sample population.  
However, an individual streamer may simply want to understand how behavior is related to individual popularity irrespective of other streamers.  This section performs streamer-centric analyses in terms of growing at a rate to reach a fixed level of success, the amount of effort streamers put in, and the effects of creating third-party accounts.

\subsection{Self-Growth Towards Partner Status}

The previous section studied models that predict whether a streamer would grow faster relative to the population. While this was useful to identify and distinguish high growth streamers, an individual streamer may simply want to improve at a steady rate in order to achieve a fixed goal.  In this case, the streamer is more interested in growing faster than a base rate.  To this end, we extended our previous temporal analysis to an outcome variable that measures ``self-growth''.  We define this based on qualifying for the Twitch Partnership Program after two years, which requires around 100 average concurrent viewers per broadcast.  Thus the base rate of growth is to gain 4 concurrent viewers per month, and the binary outcome variable measures whether this rate of growth has been achieved over a given time interval. 

\begin{figure}[h]
\centering
\includegraphics[width=.7\columnwidth]{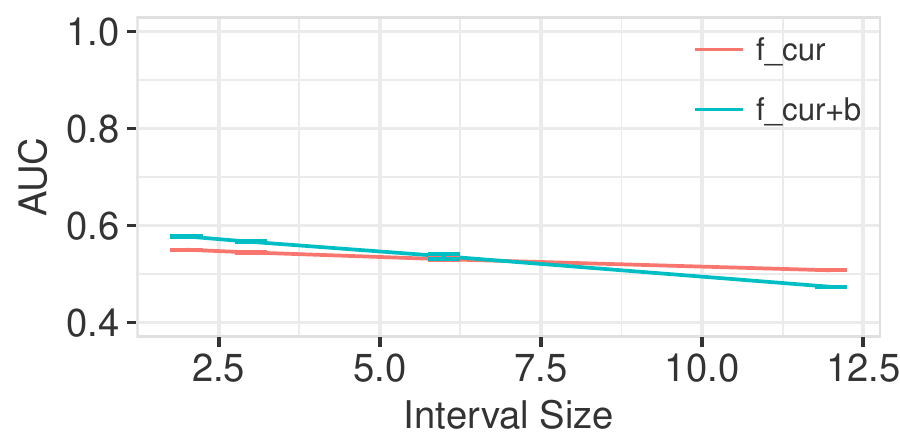}
\caption{Predicting whether streamer grows at a rate of $\ge4$ concurrent viewers per month.}
\label{f:self-growth}
\end{figure}

Figure~\ref{f:self-growth} highlights the difficulty of sustained self-growth.   Even with knowledge about streamer's past success and actions, as well as her future behavior, both the $f_{cur}$ and $f_{cur+b}$ models perform near-randomly in the short and long term.  

\subsection{Streamer Effort}

Recent media coverage~\cite{twitchnoone} suggest that many Twitch streamers spend considerable time broadcasting to no one, and that the maount of effort put in is not worth it.  Further, the preceeding study suggests that behaviors, including effort, are almost uncorrelated with the amount of concurrent viewership growth to reach Partner status.  Yet, Table~\ref{t:coefs} showed that many of highest feature coefficients were related to sheer broadcasting volume.  

To better understand these dynamics, we now study the amount of effort that streamers put into growing their popularity.   Twitch requires members of their affiliates program~\cite{twitchaffiliate} to broadcast at least 500 minutes ($8.3$hrs) per month. Thus, we use the total hours broadcasted per month as a crude measure of streamer effort.  

\begin{figure}[h]
\centering
\includegraphics[width=\columnwidth]{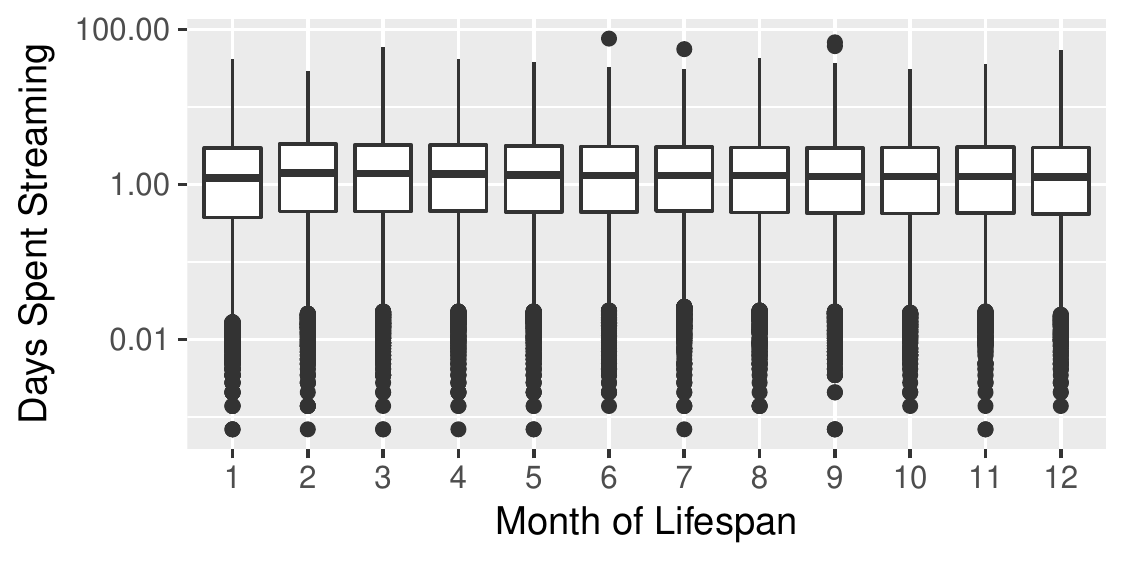}
\caption{Box Plot of the Amount of Time Spent Streaming in a Given Month}
\label{f:effort_box}
\end{figure}

Figure~\ref{f:effort_box} shows that 92\% of streamers broadcast more than the affiliates minimum.  In fact, the median streamer broadcasts for more than 24 hours per month.
We studied streamers that treat broadcasting as a full time job, as defined by broadcasting more than 40 hours per week (160hrs/month). 6\% of streamers treat Twitch as a job.  We then compared these streamers with the rest of the population by running 3 Welch Two Sample t-tests under the null hypothesis that their popularity measures are not different.  We found statistical significance for followers (effect: 5642, p-value: 1.9E-10), concurrent viewers (effect: 94.3, p-value: 8.0E-4), and cheers (effect: 171.62, p-value: 3.7E-5).  Further studies are needed to establish a causal relationship between full-time effort and success.

\begin{figure}[h]
\centering
\includegraphics[width=.8\columnwidth]{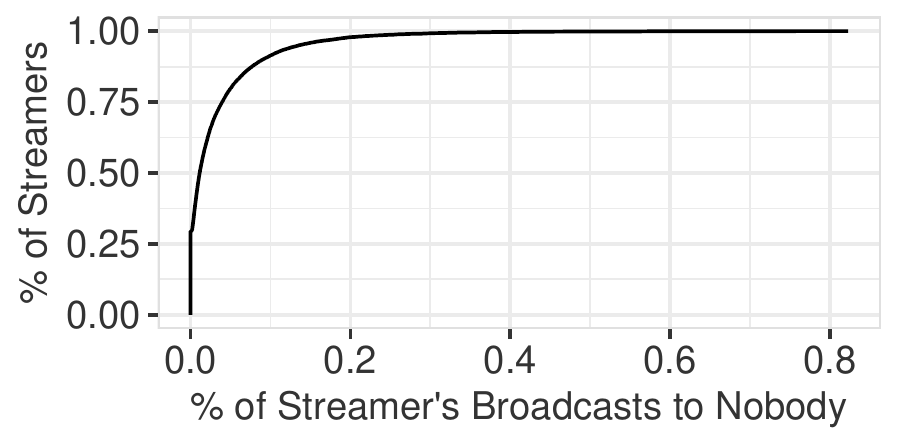}
\caption{CDF: \% of Broadcasts with 0 Viewers.  The median streamer (y-axis) has $\le1\%$ empty broadcasts (x-axis).}
\label{f:failure_density}
\end{figure}

We then studied ``failed'' streamers that broadcast to an empty audience, and found encouraging results. As shown in Figure~\ref{f:failure_density}, only 1.3\% of streamers spend more than 25\% of their broadcasts without an audience.   In fact, the majority of streamers (80\%) have less than 5\% empty broadcasts. This result suggests that it is natural to spend some amount of time broadcasting to an empty room~\cite{redditmotivated}, and most streamers that start off broadcasting to no one tend to grow out of this phase. 

\subsection{When to Create Social Media Accounts?}

We found that social media presence is correlated with follower and cumulative viewership growth. However, a streamer that is starting out without a social media presence may wonder whether creating an social media account is still worth it.  Does the timing of when an account is created have a relationship with eventual popularity?  Or are they already at a disadvantage?

\begin{figure*}[h]
  \centering
  \begin{subfigure}[t]{.32\textwidth}
    \centering
    \includegraphics[width=\columnwidth]{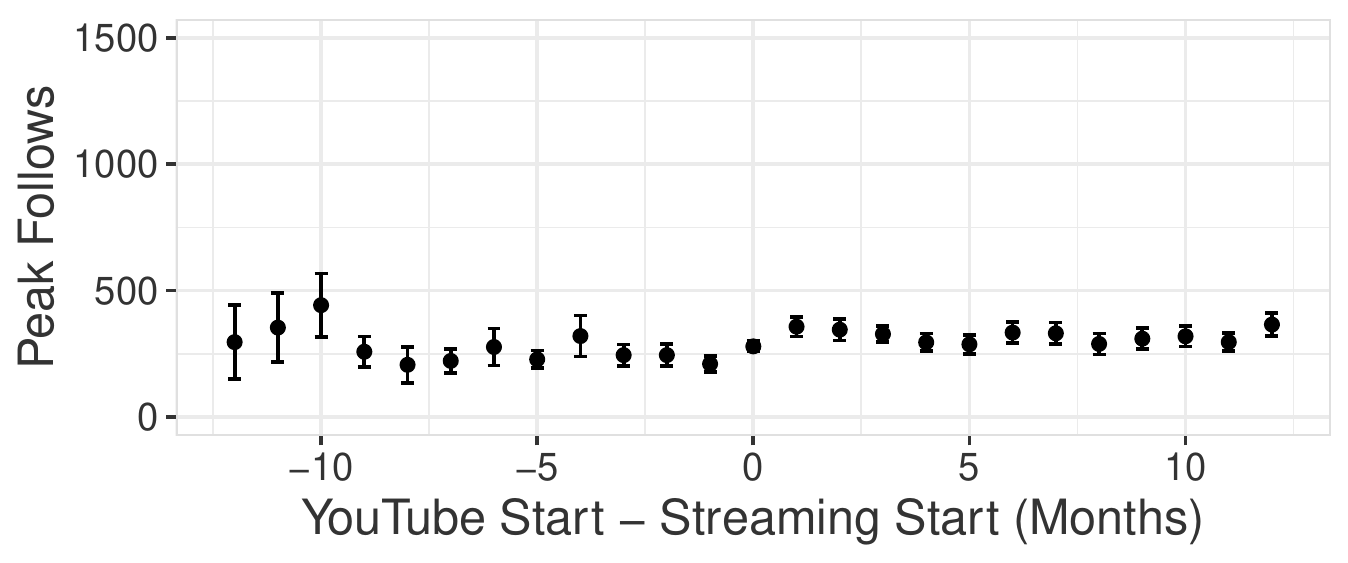}
    \caption{Peak follower count (YouTube).}
    \label{f:youtube_follows}
  \end{subfigure}
  \begin{subfigure}[t]{.32\textwidth}
    \centering
    \includegraphics[width=\columnwidth]{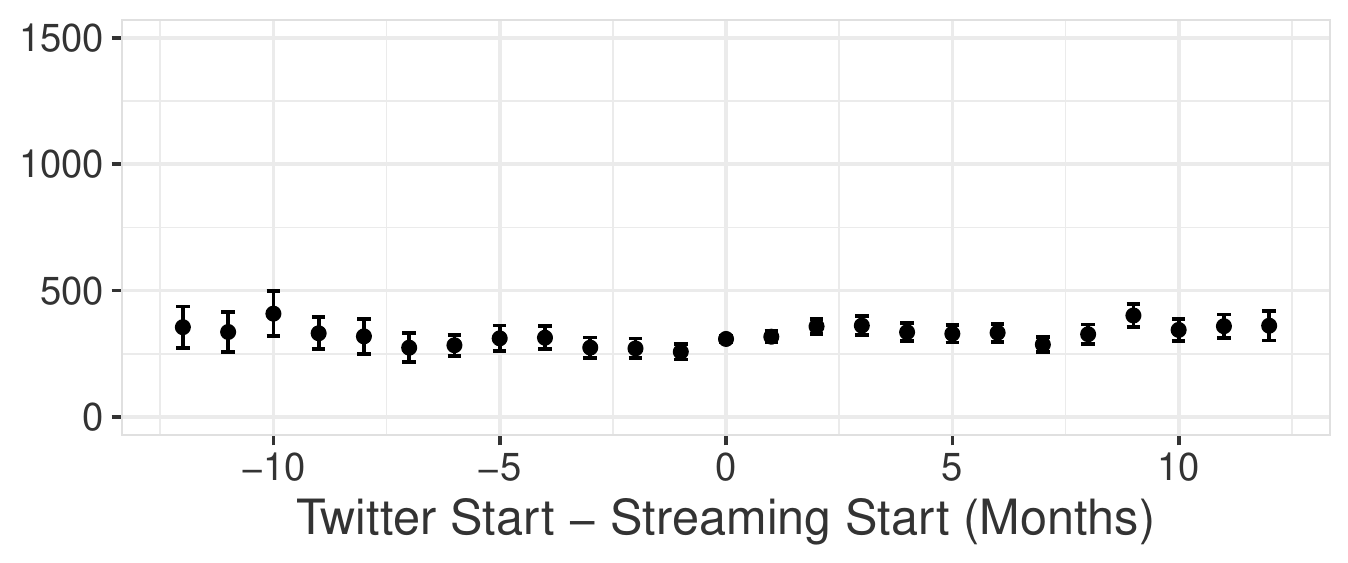}
    \caption{Peak follower count (Twitter).}
    \label{f:twitter_follows}
  \end{subfigure}
  \begin{subfigure}[t]{.32\textwidth}
    \centering
    \includegraphics[width=\columnwidth]{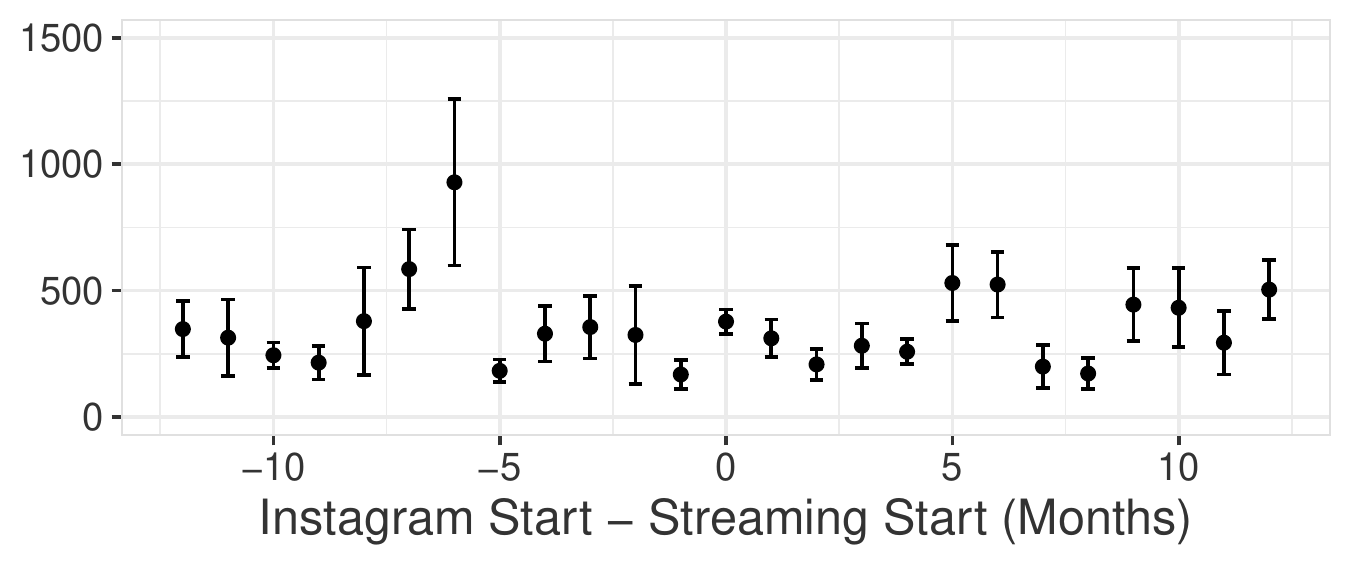}
    \caption{Peak follower count (Instagram).}
    \label{f:insta_follows}
  \end{subfigure}
  \begin{subfigure}[t]{.32\textwidth}
    \centering
    \includegraphics[width=\columnwidth]{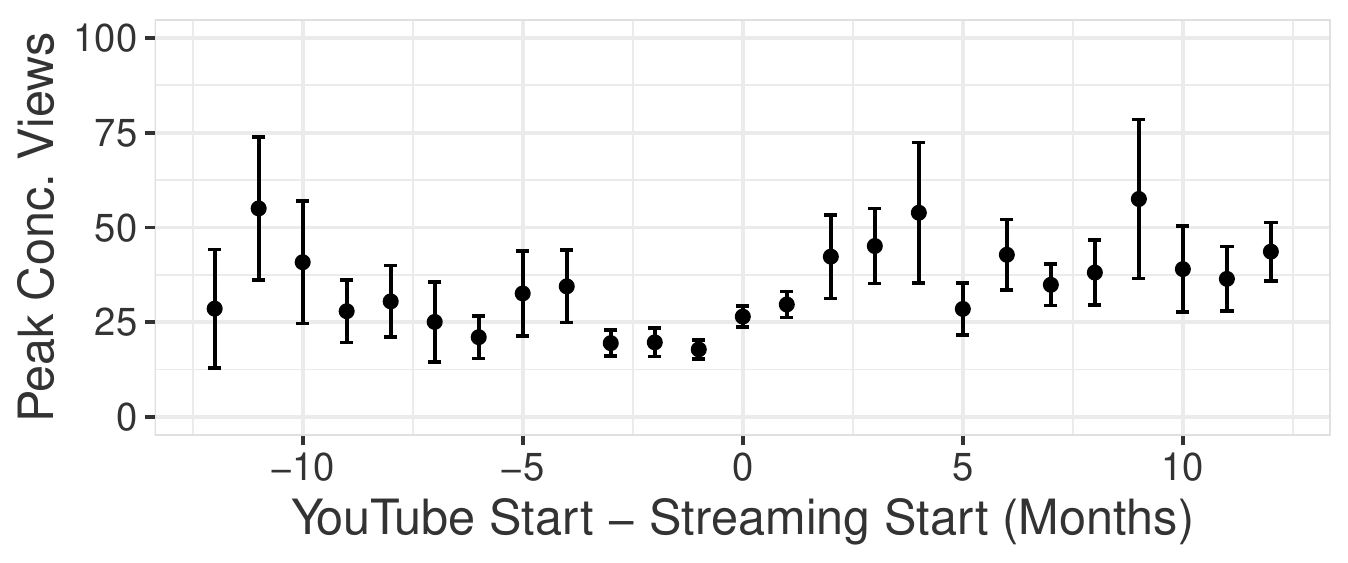}
    \caption{Peak avg. conc. views (YouTube).}
    \label{f:youtube_ccu}
  \end{subfigure}
  \begin{subfigure}[t]{.32\textwidth}
    \centering
    \includegraphics[width=\columnwidth]{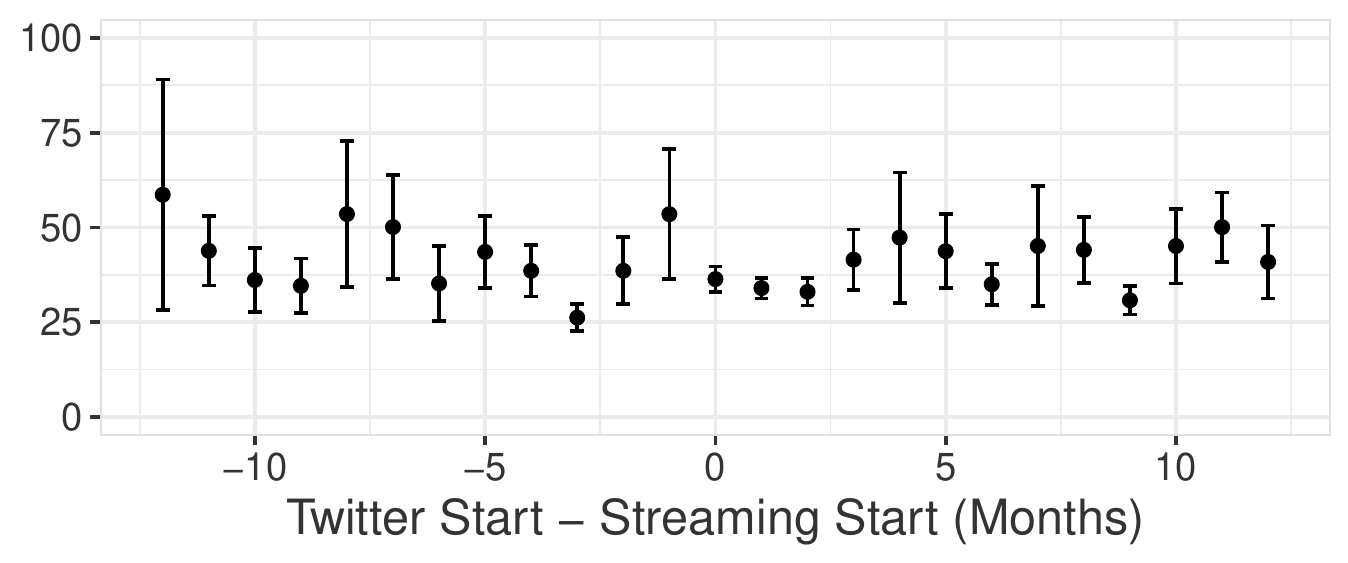}
    \caption{Peak avg. conc. views (Twitter).}
    \label{f:twitter_ccu}
  \end{subfigure}
  \begin{subfigure}[t]{.32\textwidth}
    \centering
    \includegraphics[width=\columnwidth]{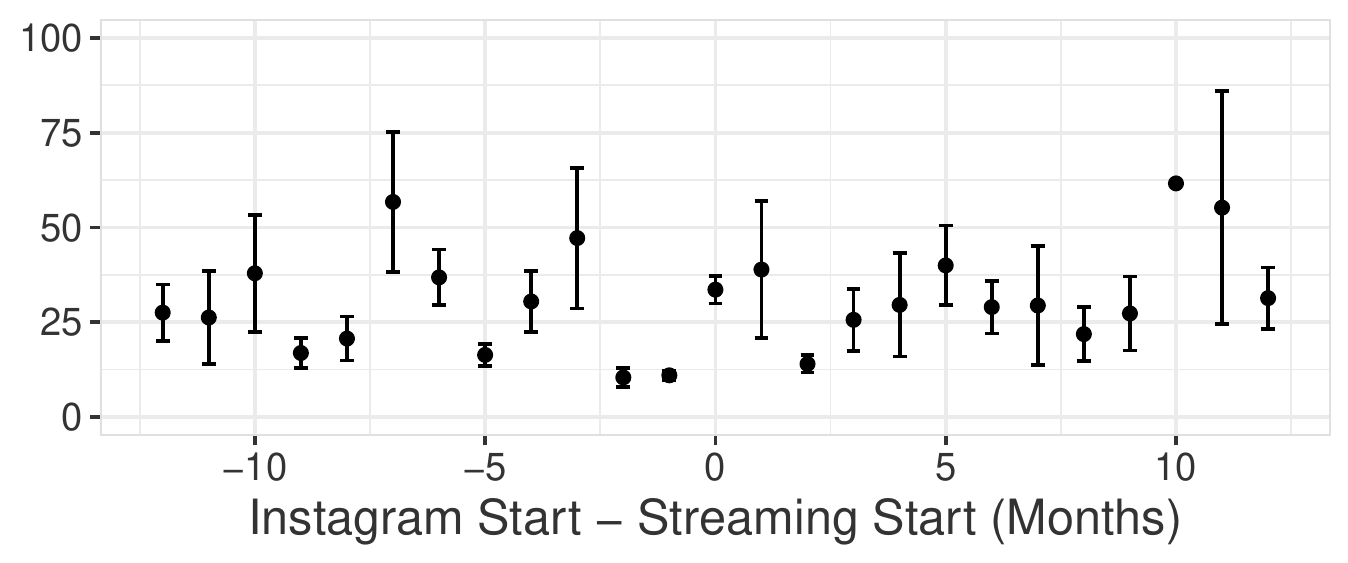}
    \caption{Peak avg. conc. views (Instagram).}
    \label{f:insta_ccu}
  \end{subfigure}
  \caption{Mean and standard error of {\it peak follower count} (top row) and
  {\it peak monthly concurrent viewership} (bottom row) across the first year of the streamer's lifespan, grouped by start month of third-party social media activity in relation to start date of streaming.}
  \label{f:soc_start}
\end{figure*}

Figure~\ref{f:soc_start} groups streamers based on when they started their YouTube, Twitter, or Instagram accounts relative to their Twitch account (x-axis).  For example, 5 in Figure~\ref{f:youtube_follows} means that the YouTube account was created 5 months after the Twitch account.  For each group, we compute the mean and standard error of the {\it peak follower count} (top row) and {\it peak monthly concurrent viewership} (bottom row) across the first year of the streamer's lifespan.

We run Welch Two Sample t-tests to compare the populations of streamers who had active social media presence before their streaming with those who created their social media accounts after. We find no statistical significance on peak follower count for each of YouTube (p-value: 0.058), Twitter (p-value: 0.128), or Instagram (p-value: 0.855). Testing again on peak concurrent views, we still find no statistical significance for YouTube (p-value: 0.09), Twitter (p-value: 0.935), and Instagram (p-value: 0.587). We see that across the different platforms and the different popularity measures,  when the social media presence begins has no effect on the peak popularity a user can achieve.   Streamers who begin broadcasting with a pre-existing social media presence do not appear to have an advantage over those that broadcast without a social media account, and even those who develop their social media presence months later.  Even though the timing does not directly affect popularity, having a social media account in itself is correlated with popularity growth.

\section{Limitations and Future Work}\label{s:discussion}

We now describe limitations and  future work.

\stitle{Beyond Modeling}  
There is a broader set of questions regarding the content creator community in general.  How does the Twitch community identify and misidentify behavioral suggestions?  Is it based on intuition?  Or survivor bias based on recommendations from successful streamers?  Furthermore, it is equally important to understand how streamers themselves choose which rules to follow---it is likely that some rules are simply easier to understand, are less resource-consuming,  or less risky.  

Models that rely on behavioral data to make predictions (e.g., Google Flu Trends~\cite{googleflu}) can directly alter and thus diverge from user behavior.   Similarly, as streamers learn about in/effective behaviors from modeling research, does their shift in behavior invalidate or alter our findings?  Revisiting these results at regular intervals may lead to interesting patterns.

\stitle{Beyond Twitch}
In this paper we have focused on Twitch, however it is unclear how our specific findings generalized to other live-streaming platforms such as YouTube-live, or more broadly, other social-media platforms.    For instance, how can a new artist, or musician, or writer, distinguish herself?  Closer to home, do academics, who produce research and publications as their primary form of content, exhibit similar characteristics?  Should academics self-promote on Twitter as well (but not too much)?   It is tempting to believe that simply producing content at a high volume and high frequency may correlate with success, however it remains to be studied. 

Despite this, we believe that our analysis methodology---to compare $F_{cur+b}$ and $F_{cur}$ on a temporal prediction task---is applicable to other studies of behavior.  Further, we believe our focus on studying the relationship between content creator behavior and long term success is both timely and important beyond Twitch.

\section{Conclusion}\label{s:conclusion}

In this paper, we study the relationship streamer behavior and popularity growth on the Twitch live-video streaming service.
We surveyed community recommended behaviors, and grouped them into 6 overarching rules.
Through careful experimental design, we seek to isolate the amount that future behavior, which streamers can control, increase the ability to predict future popularity growth. 
At the population level, we find that although behavior does not better predict how one rises through the ranks in absolute terms, it is highly correlated in identifying streamers whose relative growth is faster than the median.  From this study, we find that community recommendations are not all predictive of rapid growth, however they do not appear to harm growth either.

At the individual level, we find that it is extremely difficult to predict whether a streamer will grow at a rate to reach Twitch Partner after 2 years.  More positively, few streamers broadcast to an empty audience, creating and advertising on social media accounts is effective irrespective of when the accounts are created, and streamers that treat broadcasting as a job are more popular than the rest of the streamer population.
Ultimately, we find that the effects of user behavior on popularity growth of content-creators is a rich and deep research area, and point towards promising directions for future work in this area.
\appendix
\section{Detailed List of Features}\label{a:features}

The following is a list of features and their descriptions for each of the behavior rules used in this paper.  The features are computed with respect to a given time interval $[t, t+\delta]$.  Intervals are at least one month.

\stitle{User features}
These features were computed from attributes in the Twitch-provided dataset. They encapsulate rules 1, 2,3, and 6.
{\small
\begin{itemize}[leftmargin=*, topsep=0mm, itemsep=0mm]
  \item Broadcast Gap: The average amount of time between consecutive broadcasts.
  \item \# Broadcast: The number of broadcasts.
  \item \# Games: The average number of games per broadcast.
  \item Broadcast Len: The average length of a streamer's broadcast.
  \item \# Popular Game: The average number of popular games played per broadcast.  A game is popular if it is a top-10\% most-viewed game on Twitch.
  \item \# Days: The average number of days per week a streamer broadcasts.
  \item Sched Regularity: A measure of how consistently a streamer broadcasts on specific weekdays.  For each day of week $d$, we count the number of weeks the streamer broadcasts on that weekday $N_d$.  We then compute $\sum_{d\in [0,7]} max(N_d-1, 0)$.
  \item Unique Games: The total number of unique games played.
  \end{itemize}
  }

\stitle{Twitter features}
These features describe Rule 5, and are computed using data collected from Twitter.  If a streamer doesn't have a Twitter account, the feature is set to $0$.
{\small
\begin{itemize}[leftmargin=*, topsep=0mm, itemsep=0mm]
  \item \# Tweet: The total number of tweets.
  \item Twitter Live: Number of Twitter posts containing the word  ``live''. Streamers often advertise that they are ``going live'' before a broadcast.
  \item Tweet Before Gap: The average amount of time between a broadcast and its immediately preceeding Twitter post.
  \item Tweet After Gap: The average amount of time between the end of a broadcast and its immediately succeeding Twitter post.
  \item Twitter Adv: The number of Twitter posts with containing a Twitch URL.
  \item Tweet Len: The average character length of Twitter posts.
  \item \# Twitter Replies: The number of Twitter posts that are replies to another post.
\end{itemize}
}

\stitle{Third-Party Features}
These features are computed using the YouTube and Instagram data.  If the streamer does not have an account, the feature is set to $0$.    These features describe Rule 4.
{\small
\begin{itemize}[leftmargin=*, topsep=0mm, itemsep=0mm]
  \item \# YouTube Posts: The number of YouTube videos.
  \item YouTube Desc Len: The average length of a YouTube video's description text.
  \item YouTube Title Len: The average length of a YouTube video's title.
  \item YouTube Video Length: The average length of a YouTube video.
  \item YouTube Adv: The number of YouTube video descriptions containing a URL to Twitch.
  \item \# Instagram: The number of Instagram posts.
  \item \# Tags/Instag. Post: The average number of tags used in an Instagram post.
  \item Instagram Adv: The number of Instagram posts containing a URL to Twitch.
  \item Instagram Len: The average length of an Instagram Post.
\end{itemize}
}
\stitle{Acknowledgements}
We thank Twitch for contributing the datasets, and the Twitch data science team for helpful discussions.  
\bibliographystyle{aaai}
\bibliography{mainnew} 

\begin{thebibliography}{}

\bibitem[\protect\citeauthoryear{babybluebeam}{2016}]{lurkers}
babybluebeam.
\newblock 2016.
\newblock Let lurkers lurk.
\newblock
  \url{https://www.reddit.com/r/Twitch/comments/4t406u/let_lurkers_lurk/}.

\bibitem[\protect\citeauthoryear{Blogger}{2017}]{twitchadvice}
Blogger, T.~G.
\newblock 2017.
\newblock Ten tips to grow your creative community on twitch.
\newblock
  \url{https://blog.twitch.tv/ten-tips-to-grow-your-creative-community-on-twitch-16f3a162ff2e}.

\bibitem[\protect\citeauthoryear{Chang \bgroup et al\mbox.\egroup
  }{2014}]{chang2014specialization}
Chang, S.; Kumar, V.; Gilbert, E.; and Terveen, L.~G.
\newblock 2014.
\newblock Specialization, homophily, and gender in a social curation site:
  findings from pinterest.
\newblock In {\em CSCW}.

\bibitem[\protect\citeauthoryear{Chen}{2018}]{iceposeidon}
Chen, A.
\newblock 2018.
\newblock Ice poseidon’s lucrative, stressful life as a live streamer.
\newblock
  \url{https://www.newyorker.com/magazine/2018/07/09/ice-poseidons-lucrative-stressful-life-as-a-live-streamer}.

\bibitem[\protect\citeauthoryear{Cheng \bgroup et al\mbox.\egroup
  }{2014}]{cheng2014can}
Cheng, J.; Adamic, L.; Dow, P.~A.; Kleinberg, J.~M.; and Leskovec, J.
\newblock 2014.
\newblock Can cascades be predicted?
\newblock In {\em WWW}.

\bibitem[\protect\citeauthoryear{Clark}{2017}]{getrichnewyorker}
Clark, T.
\newblock 2017.
\newblock How to get rich playing video games online.
\newblock
  \url{https://www.newyorker.com/magazine/2017/11/20/how-to-get-rich-playing-video-games-online}.

\bibitem[\protect\citeauthoryear{Ferrara, Interdonato, and
  Tagarelli}{2014}]{ferrara2014online}
Ferrara, E.; Interdonato, R.; and Tagarelli, A.
\newblock 2014.
\newblock Online popularity and topical interests through the lens of
  instagram.
\newblock In {\em HyperText}.

\bibitem[\protect\citeauthoryear{Fu \bgroup et al\mbox.\egroup
  }{2016}]{fu2016online}
Fu, X.; Passarella, A.; Quercia, D.; Sala, A.; and Strufe, T.
\newblock 2016.
\newblock Online social networks.
\newblock In {\em Computer Communications}.

\bibitem[\protect\citeauthoryear{googflu}{2013}]{googleflu}
2013.
\newblock As flu ebbs, google tracker looking way, way too high.
\newblock
  \url{http://commonhealth.legacy.wbur.org/2013/02/google-flu-tracker-wrong}.

\bibitem[\protect\citeauthoryear{Hamilton, Garretson, and
  Kerne}{2014}]{HamiltonPlay}
Hamilton, W.~A.; Garretson, O.; and Kerne, A.
\newblock 2014.
\newblock Streaming on twitch: Fostering participatory communities of play
  within live mixed media.
\newblock In {\em ACM}.

\bibitem[\protect\citeauthoryear{Hauze}{2016}]{twitchpopularitybook}
Hauze, J.
\newblock 2016.
\newblock A guide to streaming and finding success on twitch.

\bibitem[\protect\citeauthoryear{Hernandez}{2018}]{twitchnoone}
Hernandez, P.
\newblock 2018.
\newblock The twitch streamers who spend years broadcasting to no one.
\newblock
  \url{https://www.theverge.com/2018/7/16/17569520/twitch-streamers-zero-viewers-motivation-community}.

\bibitem[\protect\citeauthoryear{Hong, Dan, and
  Davison}{2011}]{hong2011predicting}
Hong, L.; Dan, O.; and Davison, B.~D.
\newblock 2011.
\newblock Predicting popular messages in twitter.
\newblock In {\em WWW}.

\bibitem[\protect\citeauthoryear{Hutto, Yardi, and
  Gilbert}{2013}]{hutto2013longitudinal}
Hutto, C.~J.; Yardi, S.; and Gilbert, E.
\newblock 2013.
\newblock A longitudinal study of follow predictors on twitter.
\newblock In {\em CHI}.

\bibitem[\protect\citeauthoryear{Kaytoue \bgroup et al\mbox.\egroup
  }{2012}]{Kaytoue2012}
Kaytoue, M.; Silva, A.; Cerf, L.; Meira, Jr., W.; and Ra\"{\i}ssi, C.
\newblock 2012.
\newblock Watch me playing, i am a professional: A first study on video game
  live streaming.
\newblock In {\em ACM}.

\bibitem[\protect\citeauthoryear{Kim \bgroup et al\mbox.\egroup
  }{2017}]{Kim2017}
Kim, S.; Han, J.; Yoo, S.; Gerla, M.; Ciampaglia, G.~L.; Mashhadi, A.; and
  Yasseri, T.
\newblock 2017.
\newblock {\em How Are Social Influencers Connected in Instagram?}
\newblock Springer International Publishing.

\bibitem[\protect\citeauthoryear{Ma, Sun, and Cong}{2013}]{ma2013predicting}
Ma, Z.; Sun, A.; and Cong, G.
\newblock 2013.
\newblock On predicting the popularity of newly emerging hashtags in twitter.
\newblock In {\em AIST}.
\newblock Wiley Online Library.

\bibitem[\protect\citeauthoryear{Maslow}{1943}]{maslow1943theory}
Maslow, A.~H.
\newblock 1943.
\newblock A theory of human motivation.
\newblock {\em Psychological review}.

\bibitem[\protect\citeauthoryear{Parkin}{2018}]{youtubeburn}
Parkin, S.
\newblock 2018.
\newblock The youtube stars heading for burnout: 'the most fun job imaginable
  became deeply bleak'.
\newblock
  \url{https://www.theguardian.com/technology/2018/sep/08/youtube-stars-burnout-fun-bleak-stressed}.

\bibitem[\protect\citeauthoryear{Perez}{2017}]{twitchguide}
Perez, S.
\newblock 2017.
\newblock Twitch's concurrent streamers grew 67\% in q3, as youtube gaming
  declined.
\newblock \url{http://tcrn.ch/2yYMTc9}.

\bibitem[\protect\citeauthoryear{Reddit}{2018}]{twitchreddit}
Reddit.
\newblock 2018.
\newblock Twitch subreddit.
\newblock \url{https://www.reddit.com/r/Twitch/}.

\bibitem[\protect\citeauthoryear{Szabo and
  Huberman}{2010}]{szabo2010predicting}
Szabo, G., and Huberman, B.~A.
\newblock 2010.
\newblock Predicting the popularity of online content.
\newblock In {\em Communications of the ACM}.
\newblock ACM.

\bibitem[\protect\citeauthoryear{Twitch}{2018}]{twitchaffiliate}
Twitch.
\newblock 2018.
\newblock "joining the affiliate program".

\bibitem[\protect\citeauthoryear{u/AlongTheDark}{2018}]{redditfulltimetwitch}
u/AlongTheDark.
\newblock 2018.
\newblock What streaming on twitch fulltime does to your life.
\newblock
  \url{https://www.reddit.com/r/videos/comments/8m4scy/what_streaming_on_twitch_fulltime_does_to_your/}.

\bibitem[\protect\citeauthoryear{u/TheOMB}{2018}]{redditmotivated}
u/TheOMB.
\newblock 2018.
\newblock How do you stay motivated?
\newblock
  \url{https://www.reddit.com/r/Twitch/comments/83h0a8/how_do_you_stay_motivated/}.

\bibitem[\protect\citeauthoryear{Weiner}{1972}]{weiner1972theories}
Weiner, B.
\newblock 1972.
\newblock Theories of motivation: From mechanism to cognition.

\bibitem[\protect\citeauthoryear{Yang and Leskovec}{2010}]{yang2010modeling}
Yang, J., and Leskovec, J.
\newblock 2010.
\newblock Modeling information diffusion in implicit networks.
\newblock In {\em ICDM}.

\end{thebibliography}

\end{document}